# Exotic Fluids Matching the Stress-Energy Tensor of Alcubierre Warp Drive Spacetimes

Willie Béatrix-Drouhet

146 boulevard Vincent Auriol 75013 Paris, France

E-mail: drouhet@gmail.com

Abstract: To gain more insights about what could be the source of the Alcubierre warp drive, a technique designed to find an orthogonal basis from the metric expression is used. It is deduced that exotic fluids can produce the same stress-energy tensor as the Alcubierre warp drive. For different configurations of warp acceleration and velocity, constraints on density, pressure, total positive mass, total negative mass, absolute value of the equation of state as well as null-energy condition violation are provided. The Alcubierre warp drive Einstein tensor expression and the optimization of fluid properties result in power laws between various physical quantities and warp velocity. The investigated setups exhibit non-exotic equation of states for warp velocities smaller than $0.004\,c$ and seem able to sustain warp velocities up to $2\,200\,c$ with exotic equations of state.

## 1. Context and Introduction

In the context of general relativity, the Alcubierre warp drive is a spacetime configuration that theoretically would enable an object to overcome the speed of light limitation. It was first introduced by Ref. [1] and named a "warp drive" as a reference to science fiction.

However one of the most important objections found against the credibility of such a spacetime is that it does not obey the null-energy condition (NEC, see Refs. [1] [2] [3] [4] for more details).

Interestingly Ref. [1] found that the Alcubierre warp drive is not an empty spacetime. However very few direct attempt to match its stress-energy tensor with a general relativistic matter field has been documented. Thus many authors base calculation on the stress energy tensor itself whereas basing them on an adapted matter field could be interesting as well.

Besides fluids are a generic classical state of matter that has been described at length in the context of general relativity (see Refs. [5], [6] and references therein).

In addition, a fluid mix can withstand similar energy condition violations as the ones required by the Alcubierre warp drive, including NEC violation. An example of such violation is the "phantom energy", i.e. a perfect fluid with large negative pressure (see Refs. [7] [8] [9] [10] [11] [12], and references therein).

However the use of phantom energy should be complemented with precise physical quantities such as density, pressure, equation of state (EOS), energy condition violation, and so on, to confine their use to precisely documented physical boundaries linked to possible energy cut-offs.

Therefore, the methodology laid out in Ref. [13] is used to investigate whether and how an exotic fluid mix can match the Alcubierre warp bubble energy and matter requirement, exploring in greater details the physical properties of its content.

In addition to minimizing NEC violations, minimizing general relativistic fluids exoticity also means that the fluid EOS, i.e. the ratio of the total average pressure over density $P/\rho=w$, should be kept to values as close to zero as possible. More quantitatively, EOS with absolute values larger than 1 are exotic and ultra relativistic normal matter exhibits EOS of $\frac{1}{3}$ (see Ref. [2] [14] [15] [16] [17]).

In short this article deals with

- finding fluid mixes that enable to match the SET of the Alcubierre warp drive,
- restraining the exoticity of such fluids.

In section 2 usual definitions concerning the Alcubierre warp drive are given as well as orthogonal bases together with the link between such bases and fluid velocity. Ways of systematically modifying the fluid solutions are further introduced as well as the objective function that is minimized.

In section 3 the results are presented, and further discussed in section 4. Concluding remarks and a few possibilities to extend this study are given thereafter.

# 2 Methods

## 2.1 Usual Definition of the Alcubierre Warp Drive

Following Refs. [1] and [18], the square of line element writes:

$$ds^2 = g_{\alpha\beta} dx^\alpha dx^\beta = -dt^2 + (v(1-f)dt + dX)^2 + d\rho^2 + \rho^2 d\phi^2 \tag{1}$$

in geometrized units ($G=c=1$). In equation (1) has been used Einstein summation convention and usual numbering of spacetime dimensions:

$(t, X, \rho, \phi) = (x^0, x^1, x^2, x^3).$

The coordinates $(t, X, \rho, \phi)$ are time associated with cylindrical coordinates:

- $X$ is the axial coordinate,
- $\rho$, the axial distance,
- $\phi$, the azimuth.

Unless mentioned otherwise, dummy Greek indices are used for spacetime components and Latin ones for spatial dimensions.

The probe at the center of the warp drive is supposed to be a test particle of zero mass located at the origin O of coordinates $(X=0, \rho=0)$. The warp velocity is defined as $v=v(t)$, and $r=r(x^\mu)=\sqrt{X^2+\rho^2}$. Furthermore, in this article, the original definition of the $f$ radial function is used as well as $\sigma=8$ and $R=1$ (see Ref. [1]).

Besides, as mentioned in Ref. [1], $f$ exhibits significant spatial variations, thus generating potentially large tidal forces in a radially limited spatial volume close to $r=R$. As in Ref. [19] this region is named the "wall" of the warp bubble. The ingredients necessary to build this "wall" form the main focus of the present article.

## 2.2 Fully Contravariant Einstein Tensor

To analyze the content of the Alcubierre warp drive, its associated SET is computed (see Ref. [20]), i.e. the 7 independent doubly covariant Einstein tensor components $G_{\mu\nu}$ that are non zero. These results are confirmed by a detailed study with SAGE ( see Sage Notebook at Ref. [21]; at Ref. [22] some of the code used in this study is provided for replication and future work).

The expression of $G_{\mu\nu}$ only depends on

- $r$,

- $\theta$, which is the angle between the radial vector and [OX) half-line, i.e. the usual spherical polar angle,

- $v$,

- $v_t \equiv \dfrac{dv}{dt}$,

The presented Einstein tensor components are taken along cylindrical coordinates because the Alcubierre warp drive spacetime is axisymmetric, however for numerical sampling reasons expressing these components as functions of $\theta$ and $r$ is preferred to be more consistent with $f=f(r)$.

Besides to meet the Alcubierre SET with fluid mixes, the next section introduces an orthogonal basis deduced from the metric.

## 2.3 Building an Orthogonal Basis from the Alcubierre Metric

As explained by Ref. 13 , from (1) a vector basis is deduced $\{e_t, e_X, e_\rho, e_\phi\}$. Thus:

$$(e_t)^\mu = (-1, v(1-f), 0, 0),$$
$$(e_X)^\mu = (0, 1, 0, 0),$$
$$(e_\rho)^\mu = (0, 0, 1, 0),$$
$$(e_\phi)^\mu = (0, 0, 0, \frac{1}{\rho}).$$
(2)

This vector basis can be represented by the matrix $e = (e_t, e_X, e_\rho, e_\phi)$ and is orthogonal in terms of

$$\forall (\alpha, \beta), (e_\alpha)^\mu g_{\mu\nu} (e_\beta)^\nu = \eta_{\alpha\beta},$$

where $\eta_{\alpha\beta}$ is the minkowskian metric.

## 2.4 Solving Einstein Equation for a Mix of Fluids

The Einstein equation for a set of independent fluids can be written :

$$\sum_{\alpha,i} \epsilon_{i\alpha} (e_{i,\alpha})^\mu (e_{i,\alpha})_\nu = \frac{G^\mu_\nu}{8\pi}.$$
(3)

where $\epsilon_{i\alpha}$ is the $i$-th fluid's:

- density, if $\alpha = t$;
- pressure along the $\alpha$ spatial axis, if $\alpha \in \{X, \rho, \phi\}$.

The $i$-th fluid associated basis can be represented as

$$e_i = (e_{i,t}, e_{i,X}, e_{i,\rho}, e_{i,\phi}).$$
(4)

where $e_{i,t}$ gives the direction of the four-velocity of the *i*-th fluid.

From (3) and section 2.2, one can deduce that there are only 7 independent equations in (3).

Thus the object of general term $(e_{i,\alpha})^\mu (e_{i,\alpha})_\nu$ where *i* and $\alpha$ are repeated but not summed over, can be viewed as a matrix noted $[e \otimes e]$. Its dimensions are

- $4n$ columns for the combination of indices $(i,\alpha)$, where $\alpha \in \{t,X,\rho,\phi\}$,
- and 7 rows for the following couples of indices

$$(\mu\nu) \in \{(tt),(XX),(tX),(\rho\rho),(\rho X),(\rho t),(\phi\phi)\}.$$

Equation (3) can thus be expressed as the following matrix equation:

$$[e \otimes e].\epsilon = \widetilde{G},$$

where $\epsilon$ is the column vector $(\epsilon_{i\alpha})_{i\alpha}$. The effective Einstein tensor is noted $\widetilde{G}$.

Using one fluid would provide 4 unknown quantities $\{\epsilon_{1\alpha}\}_{\alpha \in \{t,X,\rho,\phi\}}$ which would result in an overdetermined system because in the expression (3) there are 7 independent non trivial equations. Overdetermined systems usually are inconsistent and thus yield theoretically approximate solutions.

From a number of fluids $n \geq 2$ obtaining exact solutions seems tenable. Thus, for the sake of simplicity only mixes of 2 and 3 fluids are considered. These configurations are respectively referred to as 2FM and 3FM in the rest of this article.

In section 2.5 the strategy to solve equation (3) is explained as well as how to build fluid associated bases, i.e. $(e_i)_i$ from expression (4).

## 2.5 Warp Drive Parameters Exploration Strategy

To solve equation (3) one needs to choose how to explore the various parameters on which $G_{\mu\nu}$ depends (see section 2.2). Warp speed acceleration $v_t$ is ascribed either of the two values $\pm 9.8 m/s^2$. $v$ values are further chosen spanning the subluminal and superluminal ranges. The $\theta$ values are chosen as sampling the interval $[0,\pi]$ by steps of $\frac{\pi}{100}$.

Similarly to Ref. [1] the explored values of $r$ are in the interval $[0.01\,R, 2.5\,R]$, linearly sampled by steps of $0.05\,R$.

The right-hand side of equation (3) is computed for fixed values of $(v, v_t)$, at each point of the following spatial grid :

$$M_j = (r_j, \theta_j)_j. \tag{5}$$

Then a fixed set of bases $\{e_i\}_i$ is picked and $\epsilon$ is computed from (3). To vary the bases $\{e_i\}_i$ so as to optimize specific properties, $e_i$, defined by (4), is expressed as a linear transformation of the $e$ basis (see expressions (2)) such that

$$e_i \equiv e_{i,y} = \Omega_y e,$$

with $y$ being a vector of real parameters. $\Omega_y$ are isometries. Their precise form is described at Ref. [22]. They represent generalized rotations and boosts adapted to curved space time.

Besides $\Omega_y$ is varied in a spatially continuous manner. Thus each coefficient from $y$ controls a linear combination of the $g_{\mu\nu}$ which itself determines parameters in the expression of $\Omega_y$. This in turn provides different versions of $\{e_{i,\alpha}\}_{i,\alpha}$.

## 2.6 Singular values of $[e \otimes e]$

Because of equation (3), $rank[e \otimes e] \leq 7$, and a necessary condition to obtain viable solutions is that $rank[e \otimes e] = 7$. If this latter condition is met, equation (3) can be solved via usual linear least square method. Thus the solution writes:

$$\epsilon = V_1 \Sigma_1^{-1} U^T \widetilde{G}, \tag{6}$$

where the block matrix representation of $[e \otimes e]$ singular value decomposition writes:

$$[e \otimes e] = U(\Sigma_1 \, 0) \begin{pmatrix} V_1^T \\ V_2^T \end{pmatrix}.$$

Because of the factor $\Sigma_1^{-1}$ in expression (6) the inverses of singular value of $[e \otimes e]$ affect the retrieved solution, namely $\epsilon$.

## 2.7 Optimization: Simulation Quality and Fluid Properties

In this section all quantities depend on $(v, v_t)$ however these dependencies are not mentioned for simplicity's sake.

How to obtain a solution of equation (3) that depends on a vector of real parameters noted $y$ is explained in last sections. However as the system is underdetermined, the solutions could further optimize a certain objective function defined from quantities that are introduced below.

The null energy condition violation ($NECV$) is introduced at each point of the grid by the following expression

$$NECV(j) = min_{i,l \in \{X,\rho,\varphi\}} \left( \{0, \varepsilon_{it} + \varepsilon_{il}\}_{M_j} \right).$$

The EOS is computed for each point $j$ and each of the $i$-th fluid:

$$EOS_{i,j} = \begin{cases} \left( \dfrac{\varepsilon_{iX} + \varepsilon_{i\rho,} + \varepsilon_{i\varphi}}{\varepsilon_{it}} \right)_{M_j}, & if \, (\varepsilon_{it})_{M_j} \neq 0 \\ 0, & elsewise \end{cases}$$

the maximum absolute value of the EOS, as

$$MAEOS(j) = max_i \left( |EOS_{i,j}|_{M_j} \right).$$

Using the above defined quantities, an objective function can be defined as follows.

$$J(j) = \begin{pmatrix} \xi_m (MAEOS(j) - 1/9)^+ \\ + \xi_{os} (overflows_j + SC_j) \\ + \xi_r (7 - rank([e \otimes e]_j)) \\ - NECV(j) \end{pmatrix}. \tag{7}$$

One alternatively minimizes the average over $j$ or the maximum over $j$ of the above objective function, by varying the vector $y$.

The combination of numerical factors,

$(\xi_m, \xi_{os}, \xi_r) = (6, 4, 6400)$,

in front of quantities entering definition (7) are heuristically chosen. One has not explored extensive variations of these parameters to find which type of solution is reached by choosing a specific parameter set.

Furthermore the objectives of expression (7), sorted from the most important to the least are as follows:

- take into account finite numerical capabilities, thus

    - the term proportional to $\xi_r$ is used to avoid rank deficiency of $[e \otimes e]$, so as to avoid computational errors,

    - the quantity named " overflows " refers to a count of all of numerical overflows that may arise in the simulation process. This ensures that the optimization is driven away from numerical overflows (see section 2.8),

- avoid a few physical conditions, notably

    - avoid producing grid cells that are more massive than a static black hole of same size (see expression (5)). To do this the SC term is defined later on in this section, referring to the Schwarzschild Condition,

    - minimize NEC violation by each of the fluids,

    - minimize the use of EOS with excessive absolute values. This is done by introducing the coefficient $\xi_m$. In this term the positive part of a real number $x$ is noted $x^+$.

One wants to avoid producing grid cells so massive that they could produce black holes if they were left isolated. Because this could contradict the spatial continuity of the Alcubierre SET. Thus considering a region of space of constant thickness located close to a plane of constant azimuth angle, the spatial volume of each grid element is computed as follows

$$V_{cell} = \Delta r \cdot \left[ r + \frac{\Delta r}{2} \right]^2 \cdot \Delta \theta^2, \qquad (8)$$

where $\Delta r$ and $\Delta \theta$ respectively are the radial and angular polar extension of the cell.

Considering that the volume of a grid cell be put into a sphere of radius:

$$R_{cell} = \sqrt[3]{\frac{V_{cell}}{\frac{4}{3}\pi}},$$

one supposes that avoiding a black hole formation requires the following condition to be fulfilled:

$$R_{cell} > R_S(M), \qquad (9)$$

where $M = V_{cell} * \sum_{i=1}^{n} \epsilon_{it}$, is the mass contained in the cell and $R_S(M) = 2M$ is the Schwarzschild radius. Thus one rewrites condition (9), and check whether it is fulfilled everywhere in the simulated map with the following condition

$$\frac{8\pi}{3} \max_j \left( R^2_{cell, M_j} \sum_{i=1}^{n} \epsilon_{it, M_j} \right) < 1. \qquad (10)$$

In expression (10) dependency on the grid cell $M_j$ is momentarily made explicit. The SC term mentioned in expression (7) is a boolean object that is 1 or 0 and represents the logical opposite of condition (10). In the rest of this article the Schwarzschild condition quantity is referred to as SCQ. This quantity is the left-hand side of expression (10).

## 2.8 Measures to Probe the Simulation

Various physical quantities are extracted from these simulations. For instance from the generalized boosts, the maximum Lorentz factors (see section 2.5) are extracted.

The following densities and pressures are also extracted

$D_{\min} = \min_{j,i} \epsilon_{it}(j),$

$D_{\max} = \max_{j,i} \epsilon_{it}(j),$

$P_{\min} = \min_{j,i} \frac{\epsilon_{iX}(j) + \epsilon_{i\rho}(j) + \epsilon_{i\phi}(j)}{3},$

$P_{\max} = \max_{j,i} \frac{\epsilon_{iX}(j) + \epsilon_{i\rho}(j) + \epsilon_{i\phi}(j)}{3}.$

as well as the following quantities:

$$-\min_j NECV(j) \qquad (11)$$

and

$$\max_{j} MAEOS(j). \qquad (12)$$

The total positive mass $M_+$ and total negative mass $M_-$ are computed. Contrary to expression (8) the volume considered to compute aggregated masses is the total volume of the cell after taking into account the rotation around the (OX) axis.

# 3 Results

This section starts from describing the properties of local quantities such as $[e \otimes e]$, and then more physical ones such as Lorentz factors, densities and pressures. $NECV$ as well as MAEOS are presented. The more global quantities combining spatial volumes and densities, such as SC as well as $M_+$, $M_-$ are presented afterward.

## 3.1 Rank of $[e \otimes e]$ and Accuracy of Retrieved Einstein Tensor

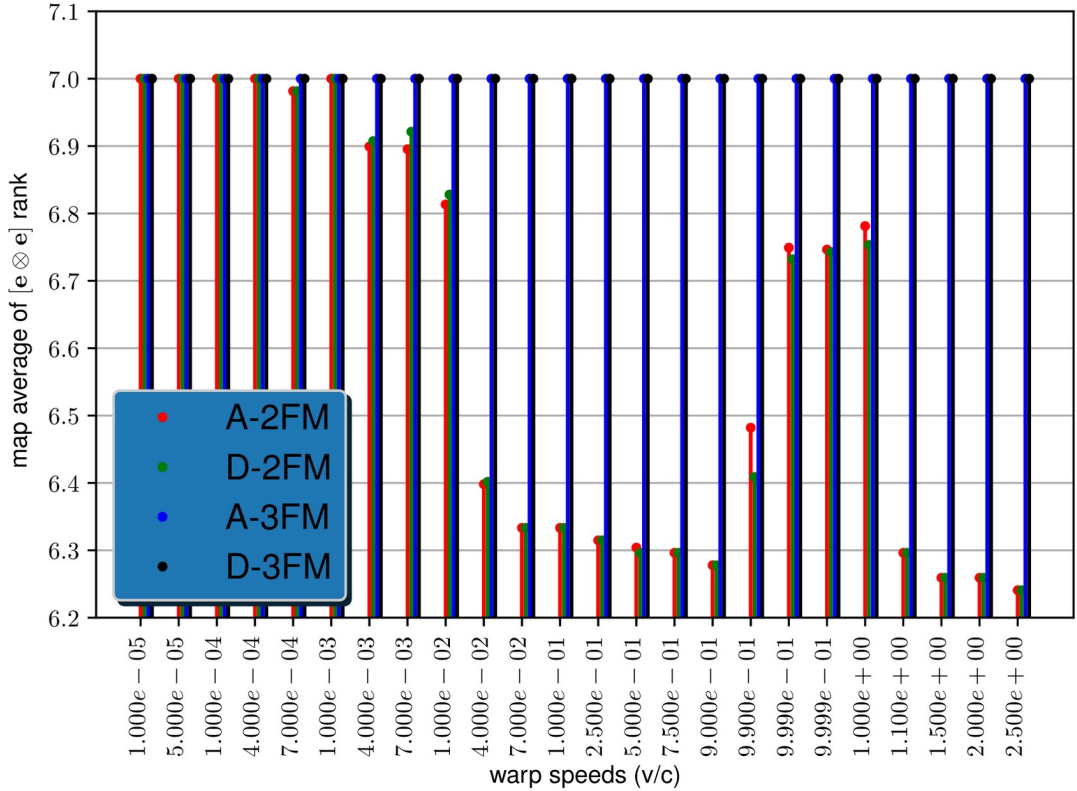

Figure 1: This figure shows variations of the average rank of [e ⊗ e] over the full range of polar coordinates (r, θ) considered. ** NB : this graph represents variations of the quantity presented above with warp acceleration, warp speed, number of fluids considered in the mix. A/D respectively means accelerating/decelerating warp drive configurations. 2FM/3FM respectively stand for stress-energy tensors matched with either 2 or 3 fluid mix.

Fig. 1 shows the grid average of ranks of $[e \otimes e]$. For consistency, only configurations fulfilling $mean_j([e \otimes e]_j)=7$ are selected and studied, the rest is discarded. Thus only a few 2FM configurations are selected and all the 3FM configurations.

## 3.2 Lorentz Factors

The maximum Lorentz factor used in these simulations is about $5e6$ found at $v=1.5c$.

## 3.3 Density and Pressure

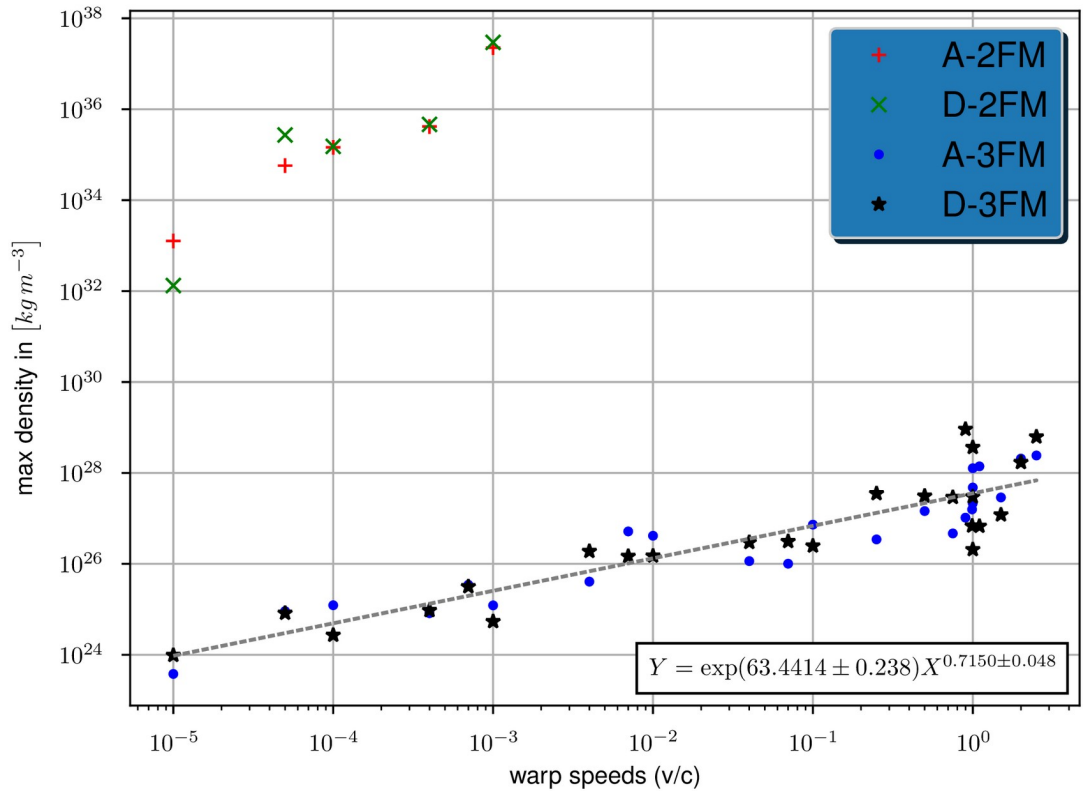

Figure 2: This figure shows log-log variations of the maximum density found throughout the simulated grid map. ** NB, see Fig. 1 for more details.

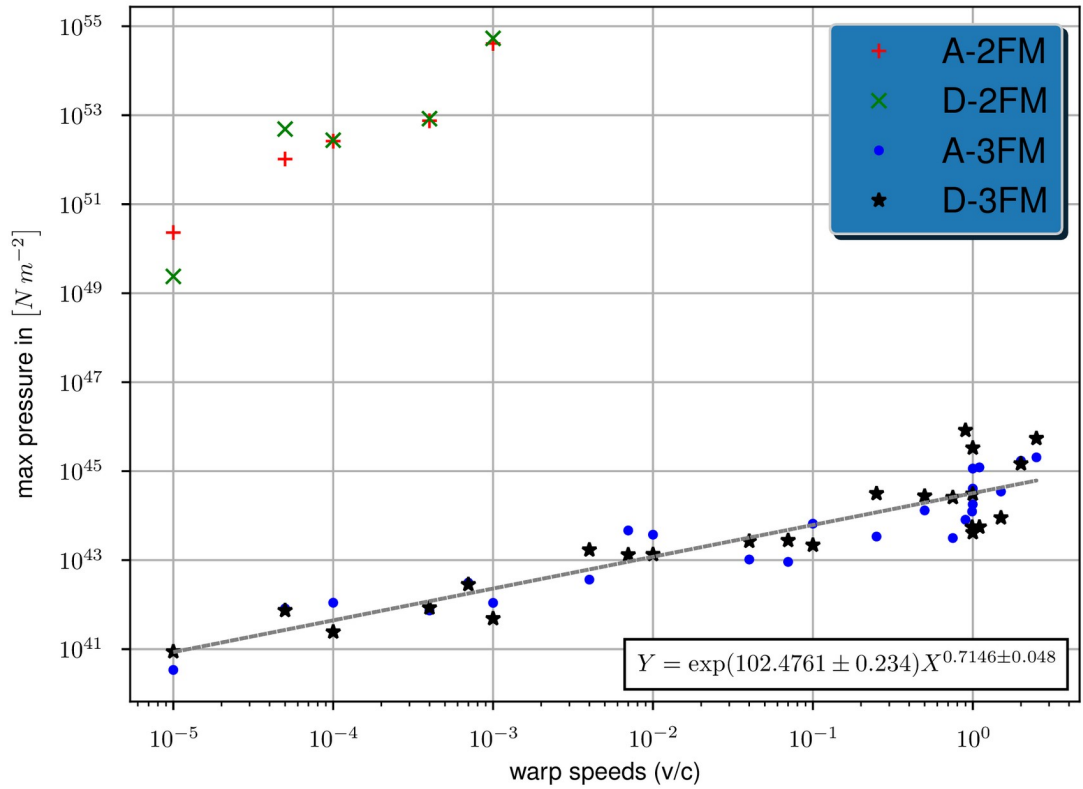

Figure 3: This figure is a log-log graph of the maximum pressure found in the full grid. ** NB, see Fig. 1 for more details.

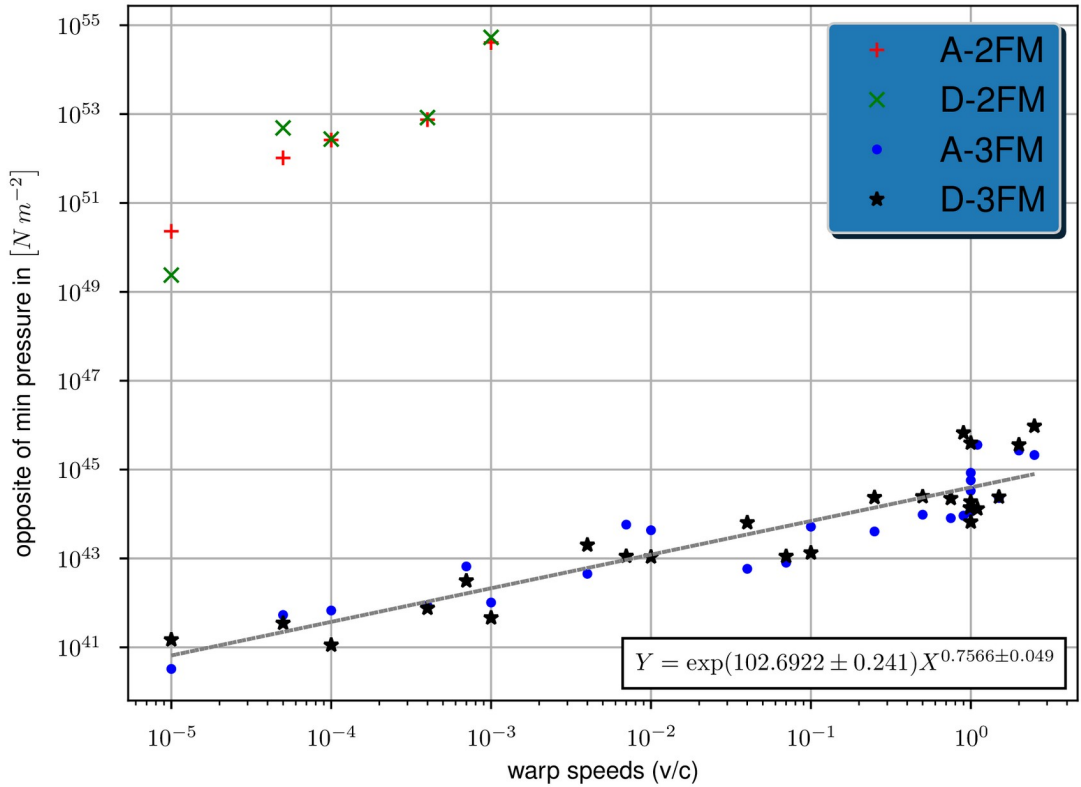

Figure 4: This figure is a log-log graph of the opposite of the minimum pressure found in the simulated grid. By opposite of the minimum pressure, we mean the maximum absolute value of the negative pressure. ** NB, see Fig. 1 for more details.

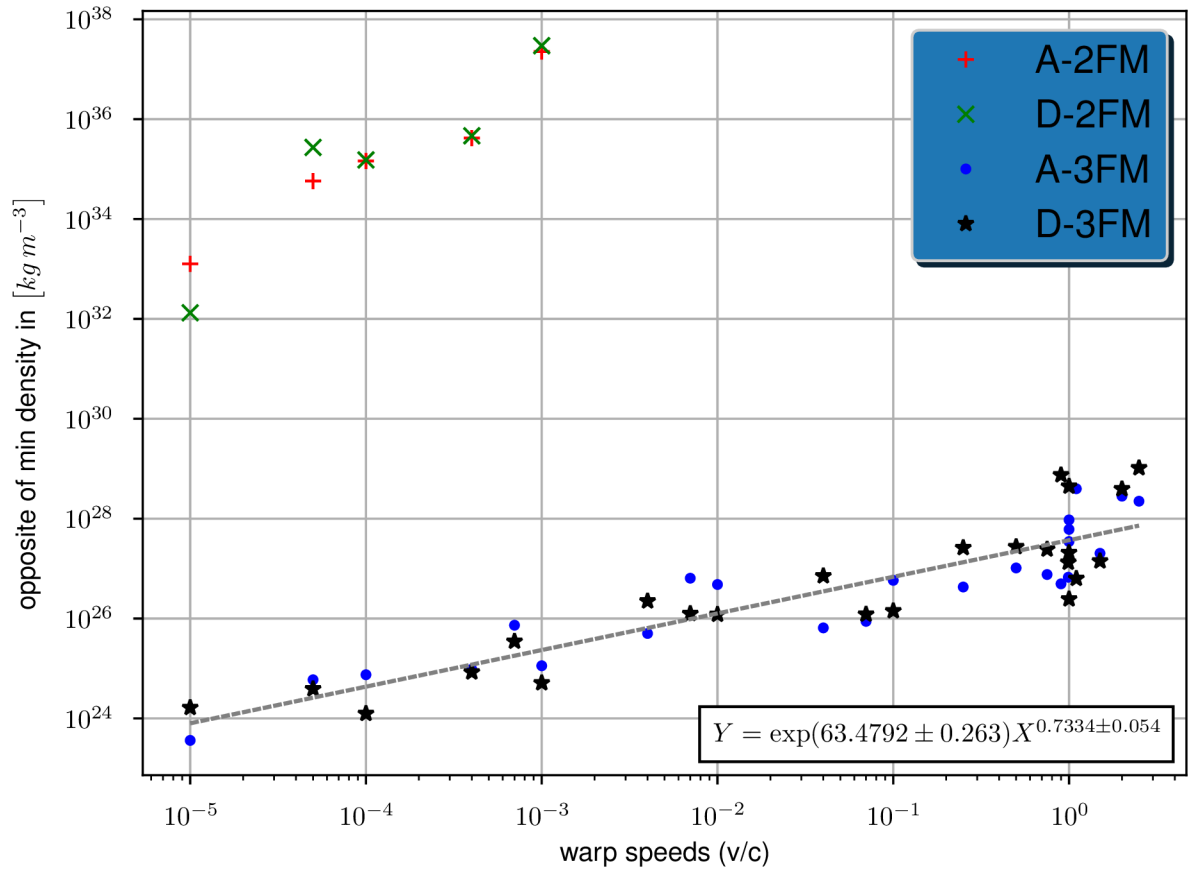

Figure 5: This figure shows log-log variations of the opposite of the minimum density that is found in the simulated grid map. By opposite of the minimum density, we mean the maximum absolute value of the negative mass density. ** NB, see Fig. 1 for more details.

Fig. 2 and 5 show the maximum density and the opposite of the minimum density. Similarly fig. 3 - 4 respectively show the maximum average pressure and the opposite of the minimum pressure. The minimum densities and minimum pressures are all found to have negative values. It is their opposite that is shown on figures. Examples of spatial distribution of fluid densities are given on figures 6-7-8-9.

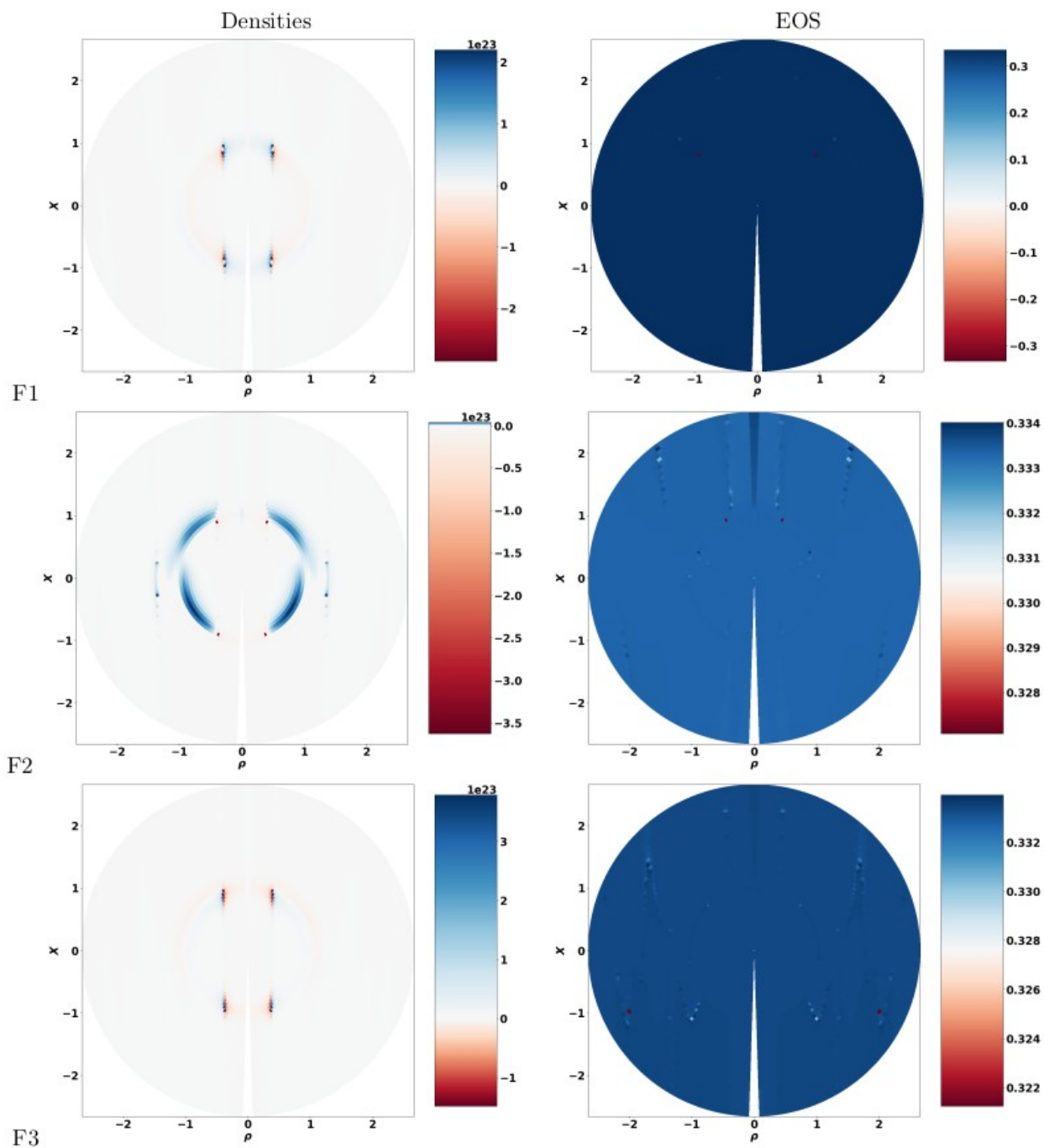

Figure 6: Densities in [ $kg\,m^{-3}$ ] and EOS obtained for fluids { F1,F2,F3 } in the 3FM configuration where v = 1e−5 , $v_t$=+1.1e-16 .

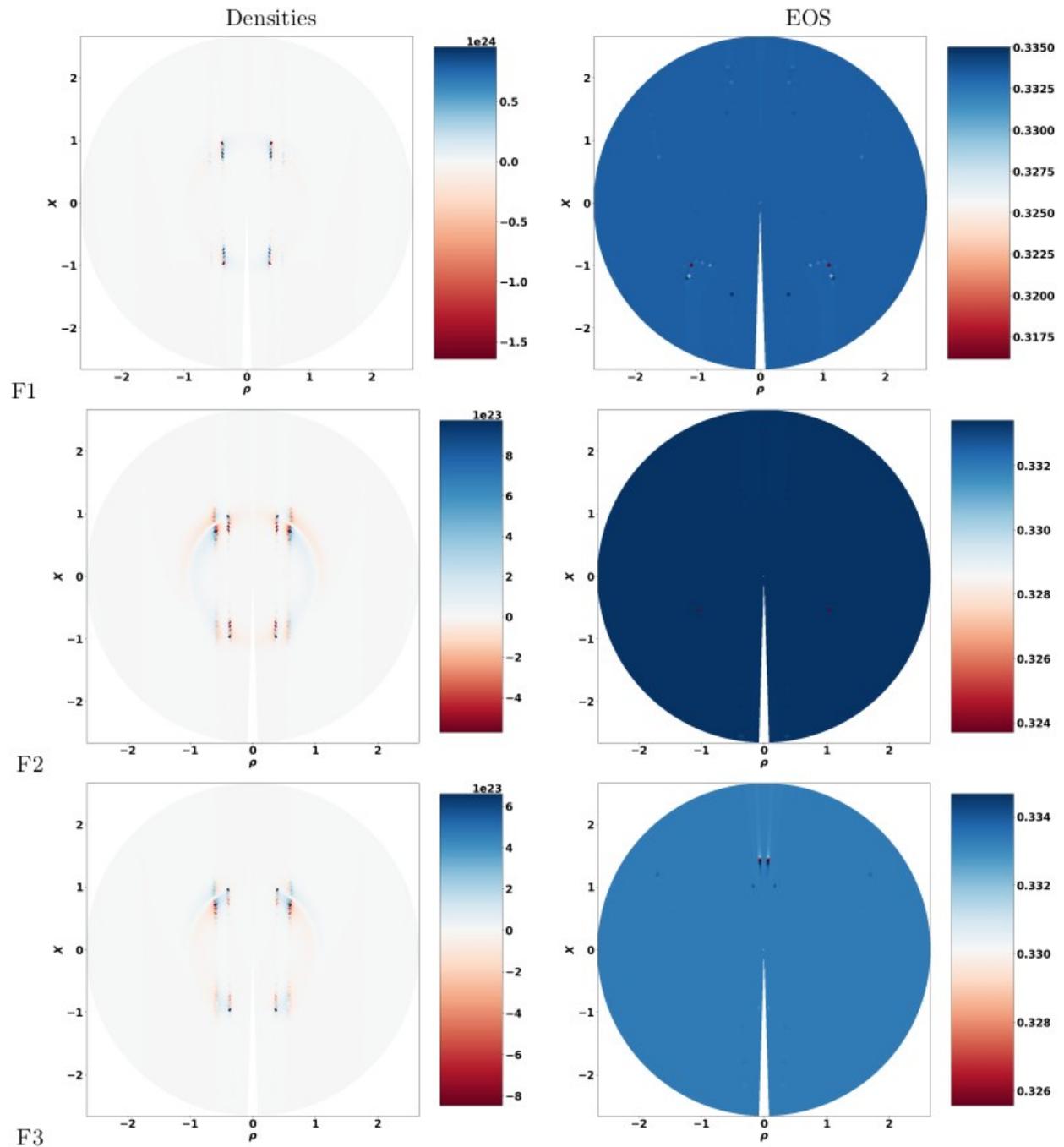

Figure 7: Densities in [$kg\,m^{-3}$] and EOS obtained for fluids { F1,F2,F3 } in the 3FM configuration where v = 1e−5 , $v_t$=−1.1e-16 .

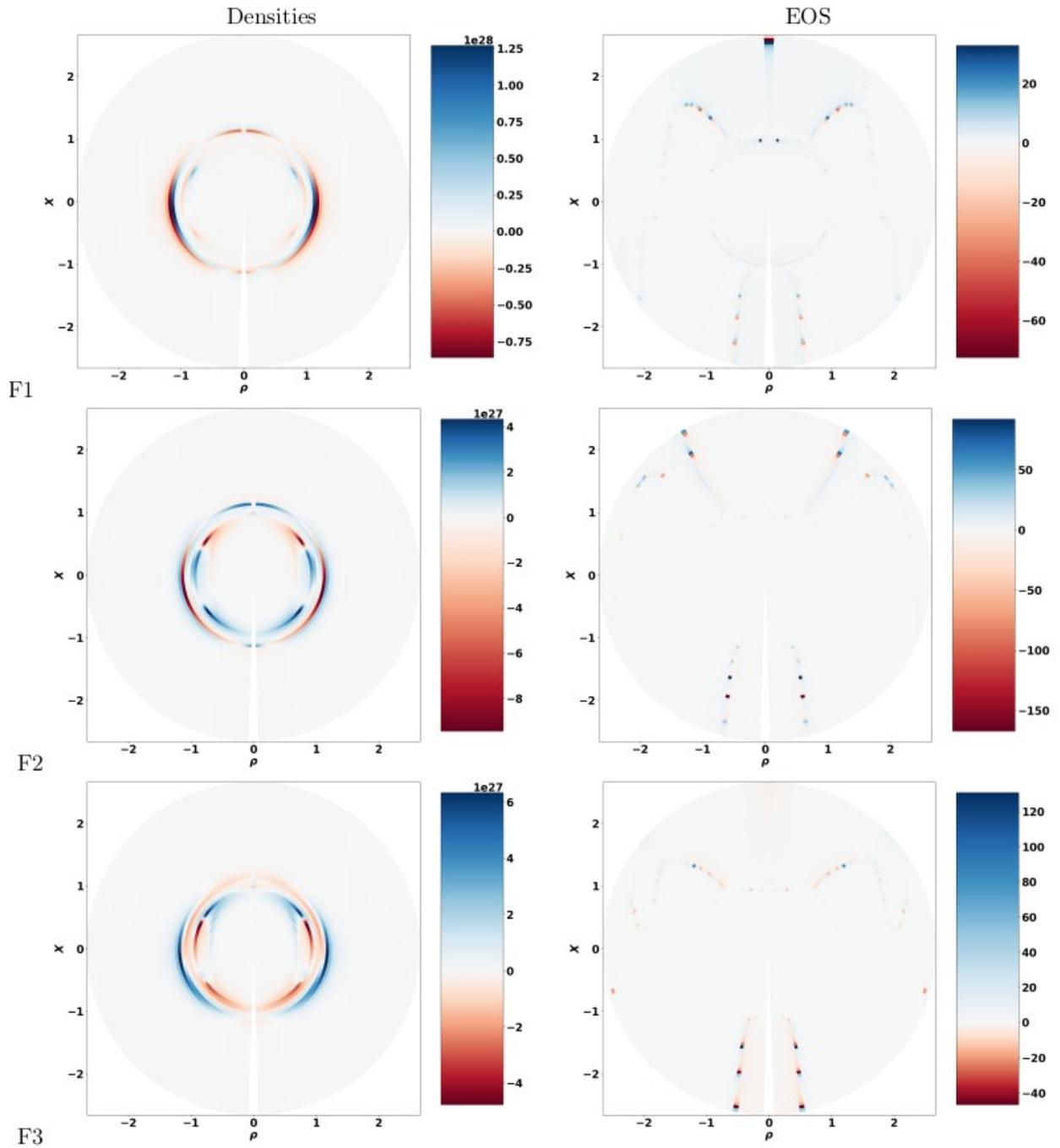

Figure 8: Densities in $[kg\,m^{-3}]$ and EOS obtained for fluids { F1,F2,F3 } in the 3FM configuration where v = 1 , $v_t$=+1.1e-16 .

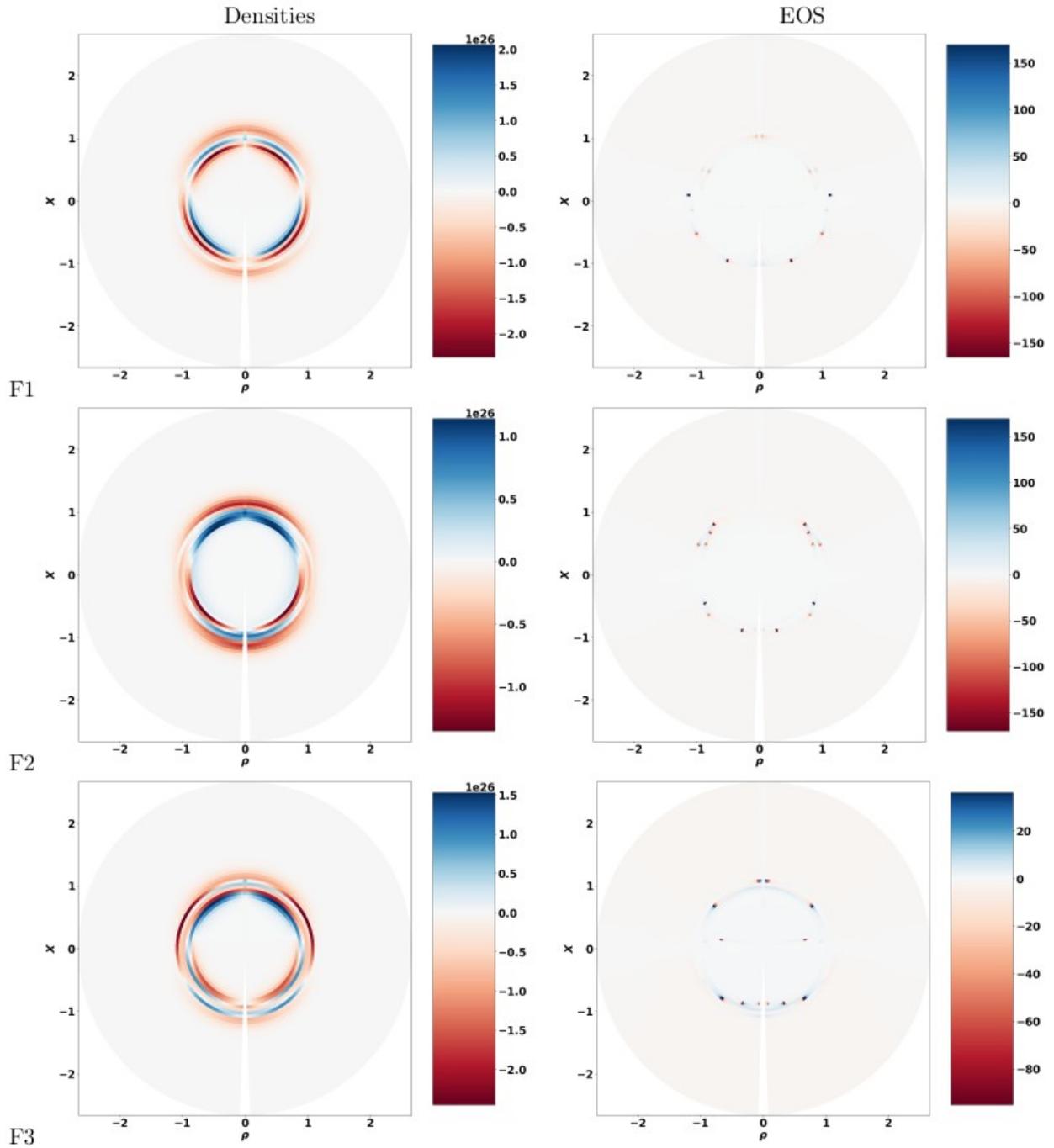

Figure 9: Densities in [$kg\,m^{-3}$] and EOS obtained for fluids { F1,F2,F3 } in the 3FM configuration where v = 1, $v_t = -1.1\text{e-}16$ .

## 3.4 Maximum Absolute Value of the Einstein tensor and Maximum Absolute Values of Densities

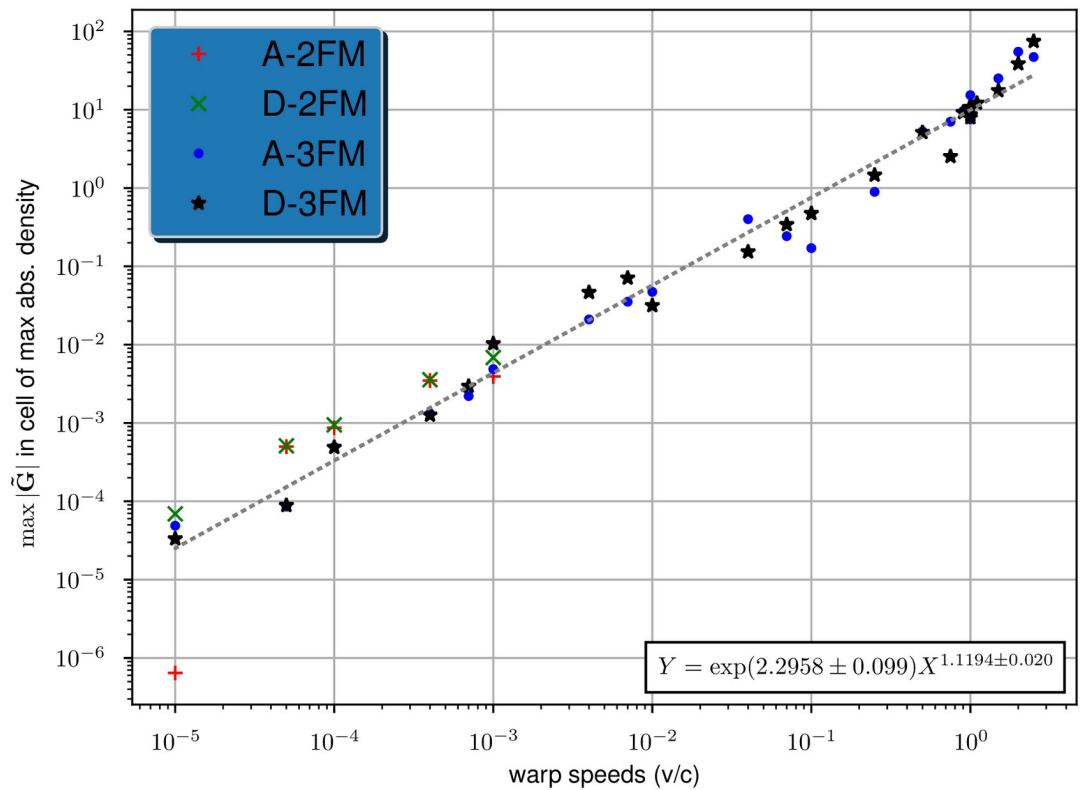

Figure 10: This figure is a log-log graph showing variations and trend (for 3FM) of the largest absolute value of the effective mixed Einstein tensor components found at the grid cell where the maximum absolute value of fluid density is also found. ** NB, see Fig 1 for more details.

From the grid where the maximum absolute value of the density is found, the maximum absolute value of the mixed-component Einstein tensor in geometrized units is extracted. This latter quantity is shown on fig. 10.

## 3.5 Null Energy Condition Violation

The quantity defined by expression (11) is represented on fig. 11. The $NECV$ of 2FM is around 10 orders of magnitude larger than for 3FM.

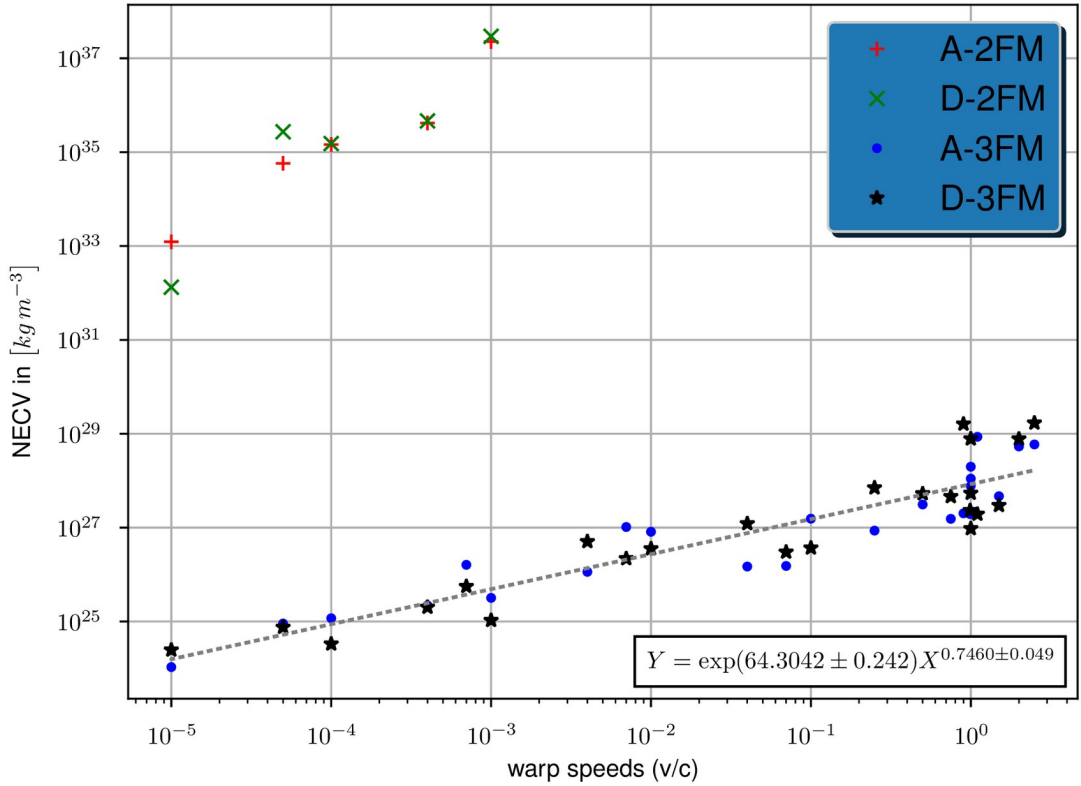

Figure 11: This figure shows variations of the opposite of the smallest NECV, found throughout the simulated grid, i.e. the quantity shown in expression (11). The (NECV) thus computed in units of density [$kg\ m^{-3}$] always is negative. The violation is more pronounced when the NECV quantity is smaller i.e. more negative. This log-log graph documents the 3FM trend between maximum NECV and v. ** NB, see Fig 1 for more details.

## 3.6 Exotic Equations of State

One also shows the maximum absolute values of the EOS, $MAEOS$ (see expression 12 ), found for 2FM and 3FM in the complete grids, on fig. 12. Examples of spatial distribution of EOS are given on figures 6-7-8-9.

From fig. 12, the only configurations enabling to achieve a $MAEOS \leq 0.35$ are 3FM with $v \leq (5e-5)c$ and the only cases of $MAEOS \leq 1$ are found for $v \leq (4e-3)c$. There is a sharp increase in the $MAEOS$, that is required to achieve the warp drive configuration, for $v \gtrsim 7e-3c$. For $v > 1e-2c$, $MAEOS > 1$.

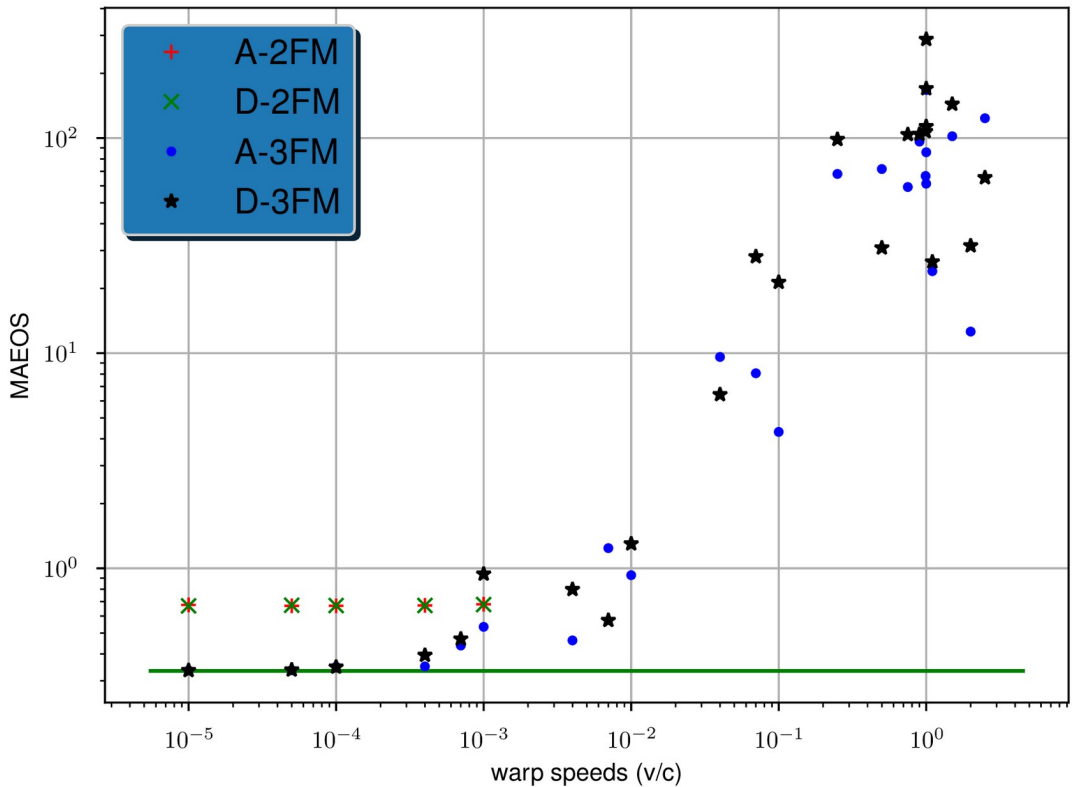

Figure 12: This figure shows variations of the maximum absolute value of the equation of state, MAEOS as expressed in 12. The equation of state is computed as the ratio between average pressure along three spatial axis and the fluid density. This log-log graph shows the trend for 3FM between MAEOS and v. The green horizontal line has an equation MAEOS = 1/3 . ** NB, see Fig 1 for more details.

## 3.7 Avoiding Black-Hole Formation

How the SCQ (see expression 10) fares is shown on fig. 13 for the different studied configurations. The reported maximum SCQ invalidates none of the results, because none of the reported configurations fulfills the SC.

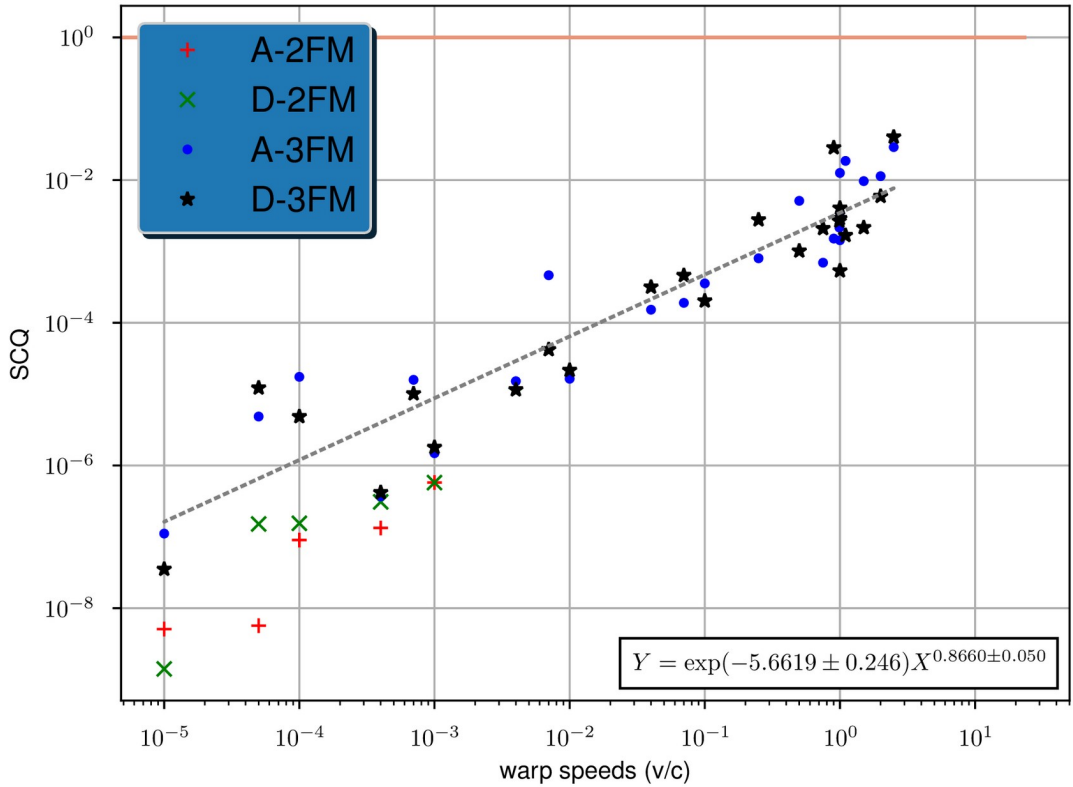

Figure 13: This figure shows variations of the left member of condition (10), named the Schwarzschild condition quantity (SCQ). SCQ is an increasing function of both volume and density. Values of SCQ larger than 1 would indicate that at least one cell of the grid is an object that would fulfill the Schwarzschild Condition and hence could develop a gravitational horizon if left isolated. This figure is a log-log graph documenting the trend between maximum SCQ and v for 3FM. ** NB, see Fig 1 for more details.

## 3.8 Total positive mass and total negative mass

Respectively fig. 14 and 15 show variations of $-M_-$ and $M_+$.

$|M_+|$ and $|M_-|$ both are smaller than 2/100 of solar mass on the studied range of $v$. Besides $M_+$ and $M_-$ are of the same order of magnitude for each $v$.

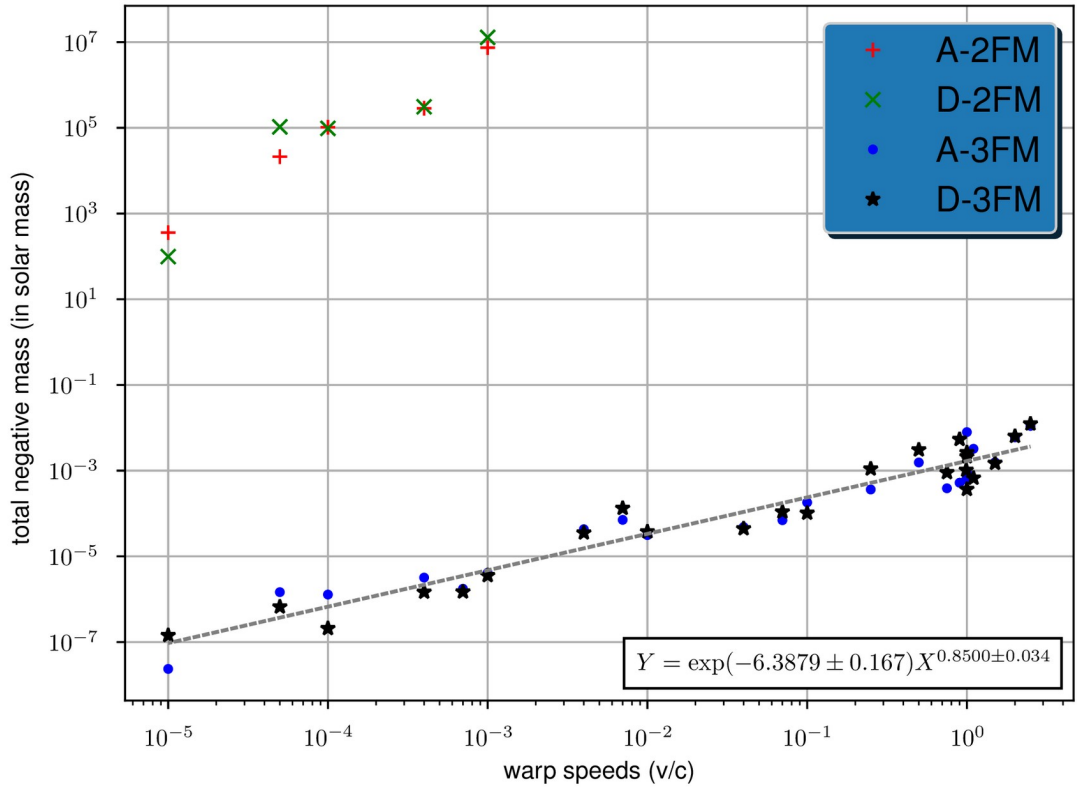

Figure 14: This log-log graph shows variations of $-M_-$ and its trend with v for 3FM. ** NB, see Fig 1 for more details.

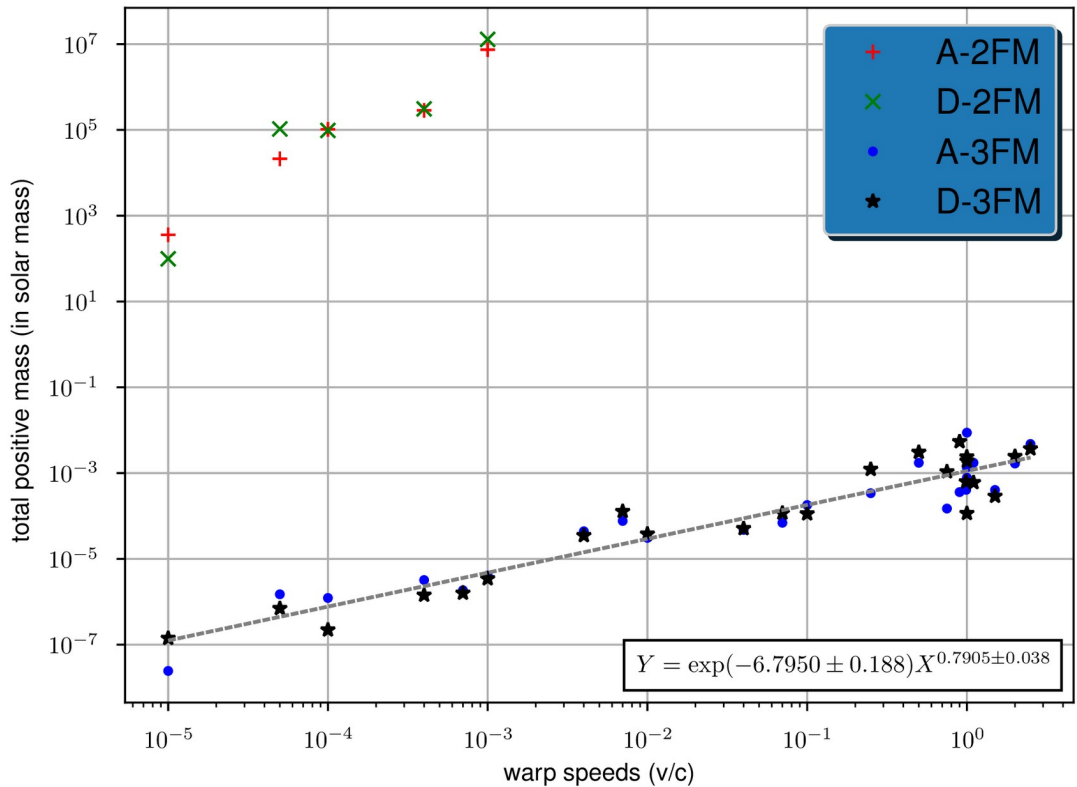

Figure 15: This figure shows variations of the $M_+$ and its trend in the case of 3FM with v. ** NB, see Fig 1 for more details.

## 3.9 Power law trends between physical quantities and *v*

Positive power law trends are found between *v* and many quantities that are extracted from the 3FM configurations. These trends are showed on figures

2-3-4-5-10-11-12-13-14-15.

## 4. Discussion

Using the commonly accepted theory of general relativity, an exotic but physically meaningful energy and matter model, namely a fluid mix is shown to match the Alcubierre SET. It is in line with comparable results found in the case of wormholes (see Ref. [13]). Besides without explicitly proving it, this work presents arguments tending to dismiss a single perfect fluid as a possible counterpart of the Alcubierre SET.

Furthermore as previously noted by many authors (see e.g. Refs. [1] and [18]), the Alcubierre warp drive violates the spacetime associated NEC. In that matter, even considering a NEC associated to each fluid component separately, the optimization pattern that is used is not sufficient to make the fluid mixes fulfill the NEC in any of the studied warp drive configuration.

Considering the EOS:

- all configurations with $v<7e-3$ use MAEOS $\leq 1$. Which is consistent with usual matter model such as scalar fields (see e.g. Ref. [15]);

- $v>1e-2$ are also simulated, but they show MAEOS $>3$. Such fluids are exotic enough to have been studied by no peer-reviewed article to date.

Concerning the large maximal Lorentz factors of about $5e6$ obtained in the simulation, one reminds that, at CERN, protons are accelerated to much smaller Lorentz factors of about $13000\,GeV/0.938\,GeV \lesssim 13800$.

The computed maximum absolute values of the mixed Einstein tensor consistent with a crude estimate based either on the warp bubble energy estimate from Ref. [19](page 3/9) or directly from $G_{\mu\nu}$ expressions. This estimate scales as $\sigma^2$ for $v=c$.

Below, aspects of the optimization process used in this study are described. First by comparing fig. 11 to fig. 5, minimum densities and $NECV$ values are similar. It means that the optimization process mainly does not finely compensate densities with pressures of opposite signs, but rather that the pressure and density are each minimized in absolute values, effectively decreasing the $NECV$.

The presented results exhibit densities (respectively pressures) necessary to match the minimal studied warp speed in the 3FM configuration that are orders of magnitude larger than the Schwarzschild stellar black hole density (respectively neutron stars typical pressures see [fig 3 of Ref. [23] ] or Ref. [24] ) but smaller than the Planck density (respectively Planck pressure). No known type of matter exhibit such characteristics. This corroborates criticism of the high energy density of warp drives by Ref. [25] quoted by Ref.[26].

Furthermore the values of maximum density and pressure retrieved in the case of 2FM are much larger than the maximum density obtained for 3FM. This result obtained with 2FM is linked to smaller singular values of $[e \otimes e]$. Smaller singular values of $[e \otimes e]$

- drive up the absolute values of extreme densities and pressures (and related quantities such as NECV) from 3FM to 2FM, while

- not impacting as much indicators which include in their definitions a competing mix of all fluids such as MAEOS, $M_+$, $M_-$ and SCQ.

Concerning $M_+$ and $M_-$, the Alcubierre metric converges toward Minkowski spacetime when $r \to \infty$, thus there is no long range effect that would be caused by an Alcubierre bubble passing through empty space time. This precludes Cerenkov radiation being created outside of the warp bubble, even if $|M_+| \neq |M_-|$.

Furthermore a few figures exhibit power law trends with respect to $v$. Typically from fig. 10 and equality (6) it seems likely that these trends originate from the links between

- $G_{\mu\nu}$ and $v$,

- the singular values of $\Sigma_1$ and $v$, i.e. how the optimization adapts differently to 3FM and 2FM configurations.

Turning to the SCQ (see expression (10)), even though this only is an informational boundary, if the 3FM power law trend of the SCQ (see fig. 13) can be meaningfully extrapolated to larger $v$, the grid map would systematically exhibit at least one cell fulfilling SC, at $v$ larger than a threshold. Larger $v$ could not be achieved within the setup studied in the present article. Computing this maximum $v$ from the power law parameters and their respective standard deviations, yields a point estimate maximum warp velocity of $2.19e3$ in units of $c$.

# 5. Conclusion

This study confirms that it is possible to find an exotic energy and matter model matching the Alcubierre warp drive. This is the case for all the warp velocities explored, either sub or superluminal, both in the decelerating and accelerating regime.

In the studied setup no warp drive configuration enables to fulfill the NEC . Apart from this, the EOS are found exotic from a warp speed equal to or larger than $0.007\,c$. Furthermore, from this simulation, building an Alcubierre warp drive with a warp velocity equal to $c$, requires around 2/100 sun mass in positive density material and a similar amount of negative density material. These mass requirements are reduced by a factor of 1e-5 for a warp velocity of $1e-5\,c$.

The optimization process that is used and intrinsic properties of the Alcubierre warp drive, produce power laws of physical quantities with respect to warp velocities. The optimization results and hence the obtained power laws, depend on the number of fluids used in the mix. Extrapolating from those power laws, one can conclude that the studied setup seems unable to reach warp velocities larger than a few thousands of $c$.

Future work could possibly modify any of the hypotheses on which this study is based. For instance:

- matter model optimization can be carried out on each grid cell,

- the various sources of numerical approximations could be explored in greater details,

- interacting fluids could be considered, and generally more realistic matter models such as the ones described by Ref.[27] involving temperature and entropy description,

- dynamical effects could be studied, e.g. the possible time related evolution of matter field used as the warp drive source.

It would also be interesting to simulate warp drives with lighter mass energy budget such as Ref. [19] , no volume expansion as in Ref. [28] or non negative Einstein tensor temporal component described in Ref. [29] to check whether these spacetimes could be produced by concrete general relativistic matter models.

## Acknowledgement

I thank for their support my wife, my family and and more generally my community, for offering help, suggestions and encouragements. I also would like to thank the reviewers for their thoughtful comments and efforts towards improving the quality of this work.